\documentclass[pra,onecolumn,amsmath,amssymb,superscriptaddress]{revtex4-1}

\usepackage{epsfig,amsmath}
\usepackage{subfigure}
\usepackage{graphicx}
\usepackage{dcolumn}
\usepackage{stmaryrd}
\usepackage{mathrsfs}
\usepackage{pifont}
\usepackage{amsthm}
\usepackage{amssymb}
\usepackage{bm}
\usepackage{latexsym}
\usepackage[colorlinks=true,linkcolor=blue,citecolor=blue]{hyperref}
\usepackage{color}
\usepackage{epstopdf}
\usepackage{float}
\usepackage{epsfig}

\usepackage{verbatim}
\usepackage{appendix}

\def\jpa#1{{ J.\ Phys.\ A} {\bf#1}}
\def\pra#1{{ Phys.\ Rev. A\/} {\bf#1}}

\def\prd#1{{ Phys.\ Rev. D\/} {\bf#1}}
\def\prl#1{{ Phys.\ Rev.\ Lett.} {\bf#1}}

\def\pla#1{{ Phys.\ Lett. A\/} {\bf#1}}
\def\rmp#1{{ Rev. \ Mod. \ Phys.} {\bf#1}}

\begin{document}

\title{Possible implications for particle physics by quantum measurement}

\author{Xiang Lv}
\affiliation{Zhejiang Institute of Modern Physics and Department of Physics, Zhejiang University, Hangzhou, Zhejiang 310027, China}

\author{Jun Jing}
\email{jingjun@zju.edu.cn}
\affiliation{Zhejiang Province Key Laboratory of Quantum Technology and Device, Department of Physics, Zhejiang University, Hangzhou 310027, Zhejiang, China}

\newcommand{\andy}[1]{ }
\def\cH{{\cal H}}
\def\cV{{\cal V}}
\def\cU{{\cal U}}
\def\bra#1{\langle #1 |}
\def\ket#1{| #1 \rangle}

\newcommand{\beq}{\begin{equation}}
\newcommand{\eeq}{\end{equation}}
\newcommand{\barr}{\begin{eqnarray}}
\newcommand{\earr}{\end{eqnarray}}

\date{\today}

\begin{abstract}
In sharp contrast to its classical counterpart, quantum measurement plays a fundamental role in quantum mechanics and blurs the essential distinction between the measurement apparatus and the objects under investigation. An appealing and popular phenomenon in quantum measurements, termed as quantum Zeno effect (QZE), can be observed in particular subspaces selected by measurement Hamiltonian rather than in the whole Hilbert space. Here we apply the top-down Zeno mechanism to the particle physics. The basic idea of Zeno mechanism is encapsulated within a consistent, unambiguous framework casted in terms of building blocks belonging to individual Zeno subspaces separated by quantum measurement. We indeed develop an alternative insight into the properties of fundamental particles, but not intend to challenge the Standard Model (SM). In a unified and simple manner, our effective model allows to merge the origin of neutrino's small mass and oscillations, the hierarchy pattern for masses of electric charged fermions, the color confinement, and the discretization of quantum numbers, using a perturbative theory for the dynamical quantum Zeno effect. Under various conditions or solutions for the transition amplitudes among particle eigenstates in the effective model, it is remarkable to probe some existing results that are somewhat reminiscent of SM, including: (i) neutrino oscillations with big-angle mixing and small masses emerge from the energy-momentum conservation, (ii) electrically-charged fermions have to hold masses in a hierarchy pattern due to the electric-charge conservation, (iii) color confinement and the associated asymptotic freedom can be deduced from the color-charge conservation. In addition, we make several anticipations about the basic properties for fundamental particles: (i) the total mass of neutrinos and the existence of a nearly massless neutrino (of any generation), (ii) the hierarchy mass pattern as well as the discretization in quantum numbers for the new-discovered electrically-charged fermions, (iii) the confinement and the associated asymptotic freedom for any particle containing more than two conserved charges. They are subject to the future improvements in both theory and experiments of higher-energy scales. 
\end{abstract}

\maketitle

\section{Introduction}

A fundamental question is always haunting in the cutting edge of physics: what are the particles as construction blocks of the universe? Many optional points of view have been proposed on the fundamental particles and their essential properties. A particle might be (i) a ``collapsed wave function'', (ii) a ``quantum excitation of a field'', (iii) an ``irreducible representation of a group'', (iv) a structure having ``so many layers'', (v) a ``vibrating string'', (vi) a ``deformation of the qubit ocean'', (vii) `` what we measure in detectors'', to name a few~\cite{quanta mag}. An ensuing question is which one, the basic particles, their mutual interaction, or their coupling with the measurement apparatus as well as the induced dynamics, is more ontological? Generally, quantum mechanics gives rise to the time evolution of particles, yet it is incapable to interpret the underlying principles running by the universe, until the data are obtained and analysed through measurements.

Answer these interesting and vital questions about particles is the starting motivation of the current work. In the Standard Model (SM), we describe the excitations of relativistic quantum fields that are characterized by fixed quantum numbers including mass, spin, and various charges~\cite{Lykken:2020xtx}. So far all the information we know about SM is obtained by detecting the quantum numbers of the fundamental fields. Quantum numbers and their structure determine what we can talk about the knowledge of physical particles. Our basic idea is that particles are nothing but a set of quantum numbers that can be measured by certain apparatus of finite resolution. The strong coupling described by quantum measurement will lead to the dynamical quantum Zeno effect (QZE) that gives rise to the low-energy/slow-oscillating effective models. As a popular and amazing effect arising from quantum mechanics, QZE~\cite{Facchi:2002, S. Pascazio:2003, Facchi:2007qz, Facchi:2009qz, S. Y. Zhu, J. Q. You, J. Jing, Maniscalco:2008zz, Mun J, Bernu J} means that a system's unitary time evolution from an eigenstate of the measured observable into a superposition will be inhibited in a subspace by the measurement-induced projection of the system. To put it another way, the fast oscillating components induced by the interactions among ``more fundamental'' particles are treated as classical variables that have to be smeared over short, finite periods of time determined by the measuring apparatus, then one obtains the effective particles and their structures for the quantum mechanical systems.

The idea that particles carrying elementary quantum numbers might be an emergent phenomenon is not brand-new~\cite{Laughlin}. However, to interpret the high-energy phenomena based on quantum measurement, such as the origin of the small masses and oscillations of neutrinos, discretization of quantum numbers and the mass hierarchy of charged leptons and quarks, and the color confinement, is conceptually markedly different in exploring particle physics. That constitutes our second motivation. We will see that quantum Zeno effect in the strong-coupling limit, as one of the most appealing effect by frequent measurements, determines what we can observe. And then particle physics is nothing but the product of quantum measurement. In particular, we will provide a succinct entry point for the associated physical phenomena to a remarkable framework described by low-energy effective models. 

\begin{table}[htbp]\centering
\begin{tabular}{|c|c|c|} \hline
Notation  & Physical implication \\ \hline
$H_M$ & Measurement Hamiltonian  \\ \hline
$P_n$, $Q_n\equiv\mathcal{I}-P_n$ & Eigenprojector for $H_M$ \\ \hline
$K=\{K_m, K_M, K_c\}$ & Coupling constants in the models for neutrino, electric charged fermion, and quark \\ \hline
$|\nu\rangle=\{|\nu_{\alpha}\rangle,|\nu_{\beta}\rangle,\cdots\}$ & Neutrino in the flavor representation \\ \hline
$\alpha, \beta=\{e, \mu, \tau\}$ & Indices for various flavor eigenstates \\ \hline
$|m\rangle=\{|m_i\rangle,|m_j\rangle,\cdots\}$ & Neutrino in the mass representation\\ \hline
$i,j=\{1, 2, 3\}$ & Indices for various mass eigenstates \\ \hline
$l_n=\{e, \mu, \tau\}$  & Charged leptons \\  \hline
$C_n=\{R, G, B\}$  &  Color fields \\  \hline
$c_n=\{r, g, b\}$  & Eigenvalues for various color fields \\ \hline
\end{tabular}
\caption{The notations and their physical implications discussed in this work.}
\label{notationtable}
\end{table}

The rest of the work is organized as following. In Sec.~\ref{theorem}, we will first briefly review the theory for quantum Zeno subspace and extend the perturbative theory to adapt to the non-Hermitian Hamiltonian. Then we construct a low-energy effective model in Sec.~\ref{model} based on two basic assumptions. This model is exploited to understand and anticipate some interesting high-energy phenomena in Sec.~\ref{phenomena}. All the discussions are focused on particles carrying only two quantum numbers with one being the conserved charge and another one being undetermined. Then we put forward three conditions as nontrivial solutions to preserve the laws of charge conservation. In particle physics, {\em Condition A, B, and C} are associated with the mass and oscillation for neutrinos, hierarchy pattern of masses for electric charged fermions, and color confinement for quarks, respectively. The physical interpretation and the basic logic of the entire theoretical framework will be reorganized in Sec.~\ref{discussion}. And finally we conclude the whole work in Sec.~\ref{conclusion}. Note in this work, we use natural units $\hbar=c=1$ and the unit of mass is chosen to be electron volt [eV]. The notations used throughout this work are listed in Table~\ref{notationtable} to avoid unnecessary misunderstanding.

\section{Review of a theory for quantum Zeno subspace}\label{Zeno subspace}

\subsection{The mathematical formulation}\label{theorem}

In the conventional presentation of the quantum Zeno effect, the coherent evolution between subspaces separated by the projective measurement will be inhibited completely. The reason behind the suppression of the evolution is the absence of the odd terms in time in the transition probability. An equivalent physical mechanism that ensures the conservation of probabilities within the relevant subspace hinges on the behavior of the survival probability, that leaks out of the subspace quadratically for short times. Then it is expected that the essential features of the QZE can be obtained by virtue of a strong continuous coupling. The mathematical formulation of this idea is contained in a theorem on the (large-$K$) dynamical evolution governed by a generic Hamiltonian of the type~\cite{Facchi:2002,S. Pascazio:2003,Facchi:2007qz}:
\beq\label{eq:sys+meas}
H_K=H+KH_M
\eeq
where $H$ is the ``more fundamental'' Hamiltonian of the interested physical system yet beyond direct observation, $K$ is a strong-coupling constant, and $H_M$ is the interaction Hamiltonian between the system and the apparatus with $\eta_{n}$, $P_{n\alpha}$ the eigenvalues and eigen-projectors, respectively. The eigen-equation for $H_M$ reads
\begin{equation}
H_{M}P_{n\alpha}=\eta_{n}P_{n\alpha}, \quad P_{n}\equiv\sum_{\alpha}P_{n \alpha}
\end{equation}
where $\alpha$ is the degenerate degree. Both $H$ and $H_M$ are assumed to be time-independent.

The full Hamiltonian in Eq.~(\ref{eq:sys+meas}) can be rescaled as
\begin{equation}
H_{\lambda}=\frac{1}{\lambda}H+H_M, \quad \lambda\equiv\frac{1}{K}
\end{equation}
Then $H_{\lambda}=\lambda H_{K}$. Now we denote $\widetilde{\eta}_{n \alpha}$ and $\widetilde{P}_{n\alpha}$ the perturbative eigenvalues and projections for $H_{\lambda}$. One can then obtain the following eigen-equation~\cite{S. Pascazio:2003}
\begin{equation}
H_{\lambda}\widetilde{P}_{n \alpha}=\widetilde{\eta}_{n\alpha}\widetilde{P}_{n\alpha}
\end{equation}
where
\begin{equation}
\widetilde{P}_{n \alpha}=P_{n\alpha}+\lambda P_{n\alpha}^{(1)}+\mathcal{O}\left(\lambda^{2}\right), \quad \widetilde{\eta}_{n \alpha}=\eta_{n}+\lambda\eta_{n\alpha}^{(1)}+\lambda^{2} \eta_{n \alpha}^{(2)}+\mathcal{O}\left(\lambda^{3}\right)
\end{equation}
A standard perturbative treatment with respect to the small $\lambda$ yields
\begin{equation}\label{projector}
\widetilde{P}_{n \alpha} =P_{n \alpha}+\lambda\left(\frac{Q_{n}}{a_{n}} H P_{n \alpha}+P_{n \alpha} H^{\dagger} \frac{Q_{n}}{a_{n}}\right)+\mathcal{O}\left(\lambda^{2}\right)
\end{equation}
and
\begin{equation}
\eta_{n \alpha}^{(1)} P_{n \alpha}=P_{n \alpha} H P_{n \alpha}, \quad \eta_{n \alpha}^{(2)} P_{n \alpha}=P_{n \alpha} H \frac{Q_{n}}{a_{n}} H P_{n \alpha}
\end{equation}
with
\begin{equation}\label{Qnan}
\frac{Q_{n}}{a_{n}}\equiv\frac{Q_{n}}{\eta_{n}-H_M}=\sum_{m \neq n}\frac{P_{m}}{\eta_{n}-\eta_{m}}, \quad Q_n\equiv\mathcal{I}-P_n
\end{equation}
Note the derivation in Eq.~(\ref{projector}) allows a non-Hermitian Hamiltonian for the fundamental particle: $H\neq H^{\dagger}$, which is popular for the effective Hamiltonian describing an open quantum system~\cite{OQS1,OQS2,OQS3,OQS4}. From now on, we set $\alpha=1$ by only considering the nondegenerate situation. While our theory can be further extended to the degenerate situation, e.g., the particles and their anti-particles are degenerate in mass.

Then with the expression for $H_K$, the perturbative time-evolution operator reads
\beq\label{fullU}
U_{K}(t)=U_{\rm{ad},K}(t)+\frac{1}{K}U_{\rm{na},K}(t)
\eeq
where
\beq\label{eq: theorem}
U_{\rm{ad},K}(t)=\exp\left\{-i\left[K H_M+\sum_{n}P_{n}H P_{n}+\frac{1}{K}\sum_{n}P_{n}H\frac{Q_{n}}{a_{n}}HP_{n}+\mathcal{O}\left(K^{-2}\right)\right]t\right\}
\eeq
is the diagonal and transitionless component and
\beq\label{nonadiabatic}
U_{\rm{na},K}(t)=\left(\sum_n\frac{Q_{n}}{a_{n}} H P_{n}\right)U_{\rm{ad}, K}+U_{\rm{ad}, K}\left(\sum_n P_{n } H^{\dagger} \frac{Q_{n}}{a_{n}}\right)+\mathcal{O}\left(K^{-1}\right)
\eeq
is the off-diagonal correction. Under the Hermitian situation $H=H^{\dagger}$, Eq.~(\ref{nonadiabatic}) will be reduced to the result in Ref.~\cite{S. Pascazio:2003}
\beq
U_{\rm{na},K}(t)=\left[\sum_{n}\frac{Q_{n}}{a_{n}}HP_{n}, U_{\rm{ad}, K}(t)\right]+\mathcal{O}\left(K^{-1}\right)
\eeq
This mathematical formation generally indicates that the evolution of a quantum system is deeply impacted in an interesting way by a continuous measurement process: this system is forced to evolve in a set of Zeno subspaces rather than the total Hilbert space in the strong-coupling limit~\cite{Facchi:2002,S. Pascazio:2003}.

\subsection{Effective model for particle physics}\label{model}

The theorem in Sec.~\ref{theorem} is used to obtain the quantum Zeno subspace as well as the associated low-energy dynamics by considering the strong coupling/measurement Hamiltonian $KH_M$ and taking $K\to\infty$. It has been argued that a ``quantum measurement" is essentially an interaction of the measured object with an external system (another quantum object, or a field, or simply another degree of freedom of the system under study) playing the role of a measurement apparatus~\cite{Facchi:2002}. The quantum measurement/observation by $H_M$ on the particle system is a two-step process, in which the system wave-function collapse is followed by rendering the information of particles. Practically, Von Neumann's postulate can then be regarded as a useful shorthand notation, summarizing the final effect of the quantum measurement~\cite{Facchi:2002} and the information of particles represented by the quantum numbers such as rest mass, electric-charge, color charges, etc. The measurement can therefore be simply decomposed to a summation of projection operators, i.e., $\sum_n\eta_nP_n$, that embodies a complicated and unknown dynamical process in charge of Zeno effects~\cite{Facchi:2009qz}. The projections actually connect the unknown/invisible high-energy physics and the physical/observable low-energy physics constrained by $H_M$.

Now we extend the proceeding theorem for Zeno mechanism to probe the origin of fundamental particles and illustrate the associated high-energy physics. The terms in the full Hamiltonian $H_K$ will be endowed with an alternative interpretation in particle physics. To be more specific, we have to introduce two basic assumptions to construct the effective Hamiltonian and model.

{\em Assumption 1}: The quantum measurement governed by any underlying dynamical process for particle physics can be effectively described by $KH_M$ with $K$ the coupling-constant satisfying the eigen-equation
\begin{equation}\label{eigenfunction}
H_MP_{n}=\eta_{n}P_{n}
\end{equation}
where $P_{n}\equiv|\eta_{n}\rangle\langle\eta_{n}|$ is the projection operator constructed by the eigenstates for $H_M$ and $\eta_{n}$ is the $n$-th eigenvalue. The eigen-equation~(\ref{eigenfunction}) defines the particles in the universe, in which $P_{n}$ can be interpreted as our observation methods or detector devices under continuous development, and $|\eta_{n}\rangle$ can be regarded as the eigenstate for the particle with quantum number $\eta_{n}$. For example, if the quantum numbers to be observed are $e$ and $m_e$, then $|\eta_{n}\rangle$ is the physical state for an electron. It should be emphasised that in our theoretical framework, the observation methods one can use or the detectors one can build is fully determined by Eq.~(\ref{eigenfunction}), which is subject to the laws in the low-energy physics, although the upper-bound of the so-called low-energy scale is continuously growing with the technique development.

In the physical picture based on {\em Assumption 1}, the measurement apparatus can be viewed as ``observers" and its observation on quantum system producing quantum numbers that are used to identify various fundamental particles. While a concrete quantum measurement described by $H_M$ generates a complicated and unknown dynamical process, its consequences can be partially treated as quantum Zeno effects. In this picture, Zeno effect/mechanism due to the quantum measurement gives rise to the collapse of the system wave-function, then the residue information about the observable fundamental particles is collected to construct the theoretical framework of Standard Model. The preceding two-step process somehow mimics the interpretation about the disappearance of the fringes in Young's double-slit interference experiment when one tries to find/measure which slit the photon travels through. That is to say, the wave-function of the photon collapses as we get the information of the photon telling its own trajectory. Note that the ``apparatus'' here can be anything as long as its interaction with the physical systems leads to the collapse of their wave-functions. We can also argue that any quantum system, including the fundamental particle, becomes reality only after it is strongly coupled to macroscopic apparatus and afterwards we can probe its information and physical mechanism.

{\em Assumption 2}: The transitionless evolution described by Eq.~(\ref{eq: theorem}) is truncated up to the leading order of $\mathcal{O}(1/K)$. It can be demonstrated by the Taylor expansion over the exponent,
\begin{equation}
\begin{aligned}\label{eq: neutrino}
& KH_M+\sum_{n}P_{n}H P_{n}+\frac{1}{K}\sum_{n}P_{n}H\frac{Q_{n}}{a_{n}}HP_{n}+\mathcal{O}(K^{-2}) \\
=& K\left[H_M+\frac{1}{K}\sum_{n}P_{n}HP_{n}+\frac{1}{K^2}\sum_{n}P_{n}H\frac{Q_{n}}{a_{n}}HP_{n}
+\mathcal{O}\left(K^{-3}\right)\right]\\
\approx & K\left[H_M+\frac{1}{K}\sum_{n}P_{n}H P_{n}+\mathcal{O}\left(K^{-2}\right)\right]
\end{aligned}
\end{equation}
The truncation in the expansion of perturbation is naturally consistent with the resolution of the measurement apparatus. Conversely, the magnitude of coupling-constant $K$ used in our assumptions is determined by the experimental resolution. Regarding Eqs.~(\ref{eq: neutrino}) and (\ref{eq: theorem}), we can have an effective Hamiltonian
\begin{equation}\label{Heff}
H_{\rm eff}=KH_M+\sum_nP_nHP_n
\end{equation}
by $H_{\rm eff}=i\dot{U}_{\rm{ad},K}(t)U^{-1}_{\rm{ad},K}(t)$. Combining the unitary evolution operator under QZE in Eq.~(\ref{fullU}) and its transition correction in Eq.~(\ref{nonadiabatic}), it renders an effective evolution operator as
\beq\label{unitary evolution operator}
U_{\rm eff}(t) =e^{-iH_{\rm eff}t}+\frac{1}{K}\left[\left(\sum_n\frac{Q_{n}}{a_{n}} H P_{n}\right) e^{-iH_{\rm eff}t}+e^{-iH_{\rm eff}t}\left(\sum_n P_{n } H^{\dagger} \frac{Q_{n}}{a_{n}}\right)\right]
\eeq
It is not necessary to specify the expression for $H_M$ for it has no direct observable effect in the context of high-energy physics beyond the energy-scale of current experiments. To ensure the physical implication of the effective Hamiltonian~(\ref{Heff}), the diagonal terms of $H$ in the eigen-representation of $H_M$ should be real. One can check whether our assumptions are correct or not by the low-energy-model interpretation for the underlying high-energy phenomena adopted in SM.

\section{Zeno mechanism for particle physics}\label{phenomena}

Particle physics of SM is basically generated by principles of gauge symmetry. Noether's theorem or Noether's first theorem states that every differentiable symmetry of the action of a physical system has a corresponding conservation law. Also, it can be expressed that if a system has a continuous symmetry property, then there are corresponding quantities whose values are conserved in time. If the symmetry groups describing physical systems are generated in a top-down way from the strong-coupling terms in the original Hamiltonian, then they yield the associated system-symmetry or charge-conservation. In the following, we will incept particle physics by focusing on particles that carry one conserved charge and another quantum number which is not necessary to be conserved and temporarily omitting the other type of charge they might have. We can demonstrate the relations between the quantum numbers and certain interesting phenomena in particle physics by virtue of Zeno mechanism, that has been manifested in Eqs.~(\ref{Heff}) and (\ref{unitary evolution operator}).

In general, all the relevant processes which have been observed and verified up to now in particle physics described by quantum numbers $\eta_n$ for any charge can be potentially categorized into the conserved and non-conserved processes. The former has been observed in the current context of particle physics and acted as a fundamental principle, which is consistent with the effective low-energy model. The latter might imply the unknown physics with higher energy. According to Eq.~(\ref{Heff}), the charge-conservation process can be described by the survival amplitude about the transition from a charge state $|\eta_n\rangle$ prepared at the beginning to the same charge state in the end:
\begin{equation}\label{survival}
\mathcal{A}(\eta_n\to\eta_n)=\langle\eta_n|e^{-iH_{\rm eff}t}|\eta_n\rangle=\langle\eta_n|e^{-i(KH_M+\sum_{m}P_{m}HP_{m})t}|\eta_n\rangle=e^{-i(K\eta_n+E_n)t}
\end{equation}
where $\eta_n$, $|\eta_n\rangle$, and $P_n=|\eta_n\rangle\langle\eta_n|$ are respectively the eigenvalue, the eigenstate and the eigen-projector of the conserved charge, and $E_n\equiv\langle\eta_n|H|\eta_n\rangle$. Consequently, the probability for charge conservation reads
\begin{equation}\label{conservation1}
\mathcal{P}(\eta_n\to\eta_n)\equiv|\mathcal{A}(\eta_n\to\eta_n)|^2=1
\end{equation}

For the same conserved charge, the second term in Eq.~(\ref{unitary evolution operator}) then describes the currently invisible process (``unphysical'' in the current context of SM). The corresponding transition amplitude reads,
\begin{equation}\label{transitionamplitude}
\begin{aligned}
& \mathcal{A}(\eta_n \to \eta_m)\\
= & \langle\eta_n|\frac{1}{K}\left[\left(\sum_n\frac{Q_{n}}{a_{n}} H P_{n}\right)e^{-i(KH_M+\sum_{l}P_{l}HP_{l})t}+e^{-i(KH_M+\sum_{l}P_{l}HP_{l})t}\left(\sum_n P_{n } H^{\dagger} \frac{Q_{n}}{a_{n}}\right)\right]|\eta_m\rangle\\
= & \frac{1}{K(\eta_m-\eta_n)}\left[\langle\eta_n|{H}|\eta_m\rangle \mathcal{A}(\eta_m \to \eta_m) - \langle\eta_n|{H^{\dagger}}|\eta_m\rangle \mathcal{A}(\eta_n \to \eta_n)\right]
 \end{aligned}
\end{equation}
where $|\eta_n\rangle$, $|\eta_m\rangle$ represent different charge states with $\eta_n\neq\eta_m$. A nonvanishing $\mathcal{A}(\eta_n\to\eta_m)$ indicates the breakdown of charge conservation. Then to preserve the charge conservation, it is required that
\begin{equation}\label{conservation2}
\mathcal{A}(\eta_n\to\eta_m) = 0
\end{equation}
in addition to the condition in Eq.~(\ref{conservation1}).

The whole effective evolution operator $U_{\rm eff}(t)$ in Eq.~(\ref{unitary evolution operator}), that is approximately unitary containing the correction to the order of $\mathcal{O}(1/K)$, determines the transitions between different eigenstates of the same quantum number or charge. The diagonal terms preserve the conserved charges defined by $H_M$, which is subject to the disturbance of the nonvanishing off-diagonal terms in the first principle. That disturbance is then expected to be cancelled through certain mechanisms or conditions, that might lead to the emergence of certain interesting phenomena in particle physics. Intuitively, the magnitude of the matrix elements $H_{nm}\equiv\langle\eta_n|{H}|\eta_m\rangle, H^{\dagger}_{nm}\equiv\langle\eta_n|{H^{\dagger}}|\eta_m\rangle$ can be simply set as small as $\mathcal{O}(1/K)$, then the magnitude of the transition amplitude $\mathcal{A}(\eta_n\to\eta_m)$ will be automatically suppressed by $\mathcal{O}(1/K^2)$. But this scenario seems a trivial solution that will not lead to any interesting physical phenomena. On the other hand, the collapse of the wave function indicates the non-unitary process determined by the off-diagonal terms of $H$ and $H^{\dagger}$. We thus only focus on the situation where the off-diagonal elements in $H$ and $H^{\dagger}$ are significant in value.

To be specific, we discuss the particles carrying two types of quantum numbers, $\eta=\{\eta_n,\eta_m,\cdots\}$ and $\nu=\{\nu_{\alpha},\nu_{\beta},\cdots\}$, whose representations can be denoted by $\{|\eta_n\rangle,|\eta_m\rangle,\cdots\}$ and $\{|\nu_{\alpha}\rangle,|\nu_{\beta}\rangle,\cdots\}$, respectively. There are two scenarios for the two quantum representations: (1) the quantum number described by $\eta$ is conserved charge but another one described by $\nu$ is not so that the two representations can be connected by a unitary transformation; (2) both quantum numbers are conserved charges so that the two representations cannot be connected by a unitary transformation (For the sake of charge conservation, they might be coincident with each other or one of them is confined to a neutral state. We here propose $3$ heuristic sufficient conditions or solutions that might give rise to Eq.~(\ref{conservation2}) at least in the order of $\mathcal{O}(1/K^2)$, which is consistent with {\em Assumption 2}.

{\em Condition A}: For very small $\eta$'s in Eq.~(\ref{transitionamplitude}), it is straightforward to check that Eq.~(\ref{conservation2}) can not be satisfied if the two representations are the same one. Then the current representation (spanned by $|\eta_n\rangle$, $|\eta_m\rangle$, $\cdots$) is connected to the representation (spanned by $|\nu_{\alpha}\rangle$, $|\nu_{\beta}\rangle$, $\cdots$) by a nontrivial unitary transformation $V$. Consequently, the transition amplitude $\mathcal{A}(\eta_n\to\eta_m)$ can be written as $\mathcal{A}(\eta_n\to\eta_m)=\sum_{\alpha,\beta}V_{\beta m}V^{*}_{\alpha n}\mathcal{A}(\nu_{\alpha}\to\nu_{\beta})$. The requirement by Eq.~(\ref{conservation2}) now becomes $\sum_{\alpha,\beta}V_{\beta m}V^{*}_{\alpha n}\mathcal{A}(\nu_{\alpha}\to \nu_{\beta})=0$, that is the conservation law expressed in the representation spanned by $|\nu_{\alpha}\rangle$, $|\nu_{\beta}\rangle$, $\cdots$. For example, in the mass representation for neutrino, the transition amplitude $\mathcal{A}(\eta_n\to\eta_m)=0$ is ensured by the mixture of $\sum_{\alpha \beta}V_{\beta m}V^{*}_{\alpha n}\mathcal{A}(\nu_{\alpha}\to\nu_{\beta})=0$ in the flavor representation, although a single transition amplitude $\mathcal{A}(\nu_{\alpha}\to\nu_{\beta})$ may survive due to the non-conservation of the lepton number. The detailed derivation can be found in Appendix~\ref{useful equations}. And this condition might be associated with the origin of light neutral fermions' masses and oscillations (see the discussions in Sec.~\ref{neutrino}).

{\em Condition B}: If the magnitude of $\eta_n-\eta_m$ in Eq.~(\ref{transitionamplitude}) is as large as $\mathcal{O}(K)$, then the transition amplitude in Eq.~(\ref{conservation2}) will be suppressed in the order of $\mathcal{O}(1/K^2)$. Note $\eta_n-\eta_m$ is the eigenvalue gaps of some kind of conserved charges, such as mass and fermion number (quantum number in the flavor representation). If $\eta_n$, $\eta_m$, $\cdots$ are identified as the masses of electrical charged fermions, e.g., charged leptons and quarks (Note here we only consider mass and flavor and ignore the color of quarks), then the significant gap among them corresponds nicely to the hierarchy mass pattern of electrical charged fermions (see the discussion in Sec.~\ref{quantization}), as an interesting phenomenon in particle physics. And this hierarchy pattern is found to be universal for any electrical charged fermions, irrespective to their flavor and the number of their generations.

{\em Condition C}: The charge states described by $|\eta_n\rangle$ or $|\eta_m\rangle$ in the transition amplitude $\mathcal{A}(\eta_n\to\eta_m)$ are confined, which means any particles carrying this kind of charges have to be combined into a uncharged composite, such that different charge states are reduced to a neutral state. Then Eq.~(\ref{conservation2}) is strictly valid without any derivation. For example, the (color) charge in the mechanism of color-confinement can never be observed in a single state, then charge conservation is naturally preserved because uncharged composites can never break charge conservation (see the discussion in Sec.~\ref{confine}). It is consistent with our very original idea that physics is the subject and result of measurement.

\subsection{Origin of masses and oscillations for light neutral fermions (neutrinos) }\label{neutrino}

All the particles we have observed are identified by a set of quantum numbers such as electric charge, mass, spin, etc. The origin of quantum numbers can be interpreted by the effective model provided in Sec.~\ref{model}. In this section, we focus on the case by identifying $\eta=m=\{m_i, m_j,\cdots\}$ as the rest mass which is small in magnitude and $\nu=\{\nu_{\alpha},\nu_{\beta},\cdots\}$ as the lepton number which is not necessary to be preserved. Then the representation spanned by $\{|m_i\rangle$, $|m_j\rangle$, $\cdots\}$ and that spanned by $\{|\nu_{\alpha}\rangle$, $|\nu_{\beta}\rangle, \cdots\}$ are identified as the mass representation and the flavor representation, respectively. Very light neutral fermions, e.g., neutrinos, are the practical particles satisfying the above conditions.

We apply Eq.~(\ref{eigenfunction}) to interpret the origin of the mass of neutral fermions. Specifically the strong coupling-constant $K$ is written as $K_m$ and the eigen-equation in the mass representation is now rewritten as
\begin{equation}\label{eigenfunction for neutrino}
H_M P_{i}=m_{i}P_{i}, \quad P_i\equiv|m_i\rangle\langle m_i|
\end{equation}
where $m_i$ and $|m_i\rangle$ denote the mass and mass eigenstate for the $i$-th generation neutral fermion, respectively. Here we focus on the case for the $3$-generation light neutral fermions and one will see in Appendix~\ref{useful equations} that the existence of more generation neutral fermions is a straightforward generalization.

According to {\em Condition A}, where the two different representations are related by a nontrivial unitary matrix of the $3$-generation model~\cite{Akhmedov:1999uz}
\begin{equation}
\left(\begin{array}{c}
\nu_{e } \\
\nu_{\mu } \\
\nu_{\tau }
\end{array}\right)=\left(\begin{array}{ccc}
V_{e 1} & V_{e 2} & V_{e 3} \\
V_{\mu 1} & V_{\mu 2} & V_{\mu 3} \\
V_{\tau 1} & V_{\tau 2} & V_{\tau 3}
\end{array}\right)\left(\begin{array}{l}
m_{1 } \\
m_{2 } \\
m_{3 }
\end{array}\right)
\end{equation}
Also, it can be briefly expressed by
\begin{equation}\label{u-matrix}
|\nu_{\alpha}\rangle=\sum_{i}V_{\alpha i}|m_{i}\rangle
\end{equation}
where $\alpha\in\{e,\mu,\tau\}$ and $i\in\{1,2,3\}$.

Starting from Eqs.~(\ref{transitionamplitude}) and (\ref{u-matrix}), one can figure out a group of linear algebraic equations for the Hamiltonian elements corresponding to the ``more fundamental'' physics for the light neutral fermions. It follows with the energy-momentum conservation in the mass representation while demonstrates oscillations in a different representation (the flavor representation). The details of derivation and analysis for the transition amplitude in the flavor representation can be found in appendix~\ref{useful equations}. Based on the $K$-independent result found in appendix~\ref{useful equations}, we can further obtain the formula in appendix~\ref{oscillationFormula} for the probability of neutral fermion oscillations:
\begin{equation}\label{final}
P(\nu_{\alpha} \to  \nu_{\beta} )=\delta_{\alpha \beta}-4 \sum_{i>j} \Re\left(V_{\alpha i} V^{*}_{\beta i} V^{*}_{\alpha j} V_{\beta j}\right) \sin ^{2} \frac{\Delta m_{i j}^{2} L}{4 E}+2 \sum_{i>j} \Im \left(V_{\alpha i} V^{*}_{\beta i} V^{*}_{\alpha j} V_{\beta j}\right) \sin \frac{\Delta m_{i j}^{2} L}{2 E}
\end{equation}
where $L$ is the distance which the neutral fermion has propagated, $E$ is the mean energy for neutral fermion, $\Re$ ($\Im$) indicates the real (imaginary) part, and $\Delta m_{ij}^{2}\equiv m_i^2-m_j^2$. This result is exactly the same as the previous ones~\cite{Ternes:2020bvy,Esteban:2020opq,Torri:2020dec,Pascoli2019}.

Neutrinos in SM are the most interesting neutral fermions that have been found in experiments. They play a crucial role in various areas, such as subatomic physics, astrophysics and cosmology. They are the most mysterious and fascinating in all of the elementary particles puzzling physicists. It is common believed that they play a very important role in both the microscopic view of particle physics and the macroscopic view of evolution of the universe~\cite{Kang:2020ghg}. The smallness of neutrino mass is related to the existence of unexplored mass scales in particle physics with a high probability. These energy-scales are so high that their direct experimental exploration might be always beyond our reach. Neutrinos can nevertheless provide us with indirect information about their mass scales and the novel physics associated with these scales, e.g., they may even hold a clue to the general problem of the fermion mass generation~\cite{Akhmedov:1999uz}.

The discovery of neutrino oscillation or mixture demands that neutrinos should have non-zero but uniquely small masses, and consequently, the SM of particle physics is incomplete. Regarding the origin of neutrino masses and neutrino oscillations, there are three main models: see-saw type I~\cite{Minkowski:1977sc, M. Gell-Mann, T. Yanagida, R.N.Mohapatra},  see-saw type II~\cite{Magg:1980ut, J.Schechter, C.Wetterich, G.Lazarides, R.N.Mohapatra2}, and see-saw type III~\cite{Foot:1988aq, E.Ma} based on the Weinberg operator~\cite{Weinberg:1979sa} explaining both existence and smallness of neutrino masses. These seesaw mechanisms, however, cannot interpret the origins of neutrino oscillations and their big-angle mixing in the first principle.

With the results in appendices~\ref{oscillationFormula} and \ref{useful equations}, we find that (1) the oscillation among flavor eigenstates is the only consistent manner for very light neutrino preserving energy-momentum conservation, which can be described by the probability formula for oscillation in Eq.~(\ref{final}); (2) Not all of the mixing angles for neutrinos can be small, i.e., there is at least a big-angle mixing in the oscillations; (3) Using the experimental data for the neutrino physics~\cite{SinghKoranga:2020jzh}, the total mass for the $3$-generation neutrinos $\sum m_i$ is evaluated in our model as $0.059$ eV for the normal ordering ($\Delta m_{31}^{2}>0$) or $0.101$ eV for the inverse ordering ($\Delta m_{31}^{2}<0$) and there exists a nearly massless neutrino. Note the constraining cosmological upper bound to date on $\sum m_{i}$ can be obtained by combining the cosmological observations: Cosmic Microwave Background (CMB) with different large scale structure observations, which ranges from $\sum m_{i}<0.12$ eV to $\sum m_{i}<0.15$ eV at $95\%$ confidence level (CL)~\cite{M.Lattanzi, S.Vagnozzi, E.Giusarma, M.Gerbino, A.J.Cuesta, N.Palanque-Delabrouille, E.Di Valentino}. Thus our findings are consistent with the existing theoretical and experimental results.

\subsection{Discretization of quantum numbers}\label{quantization}

As a class of important phenomena which are closely related to the high-energy physics, the characteristics of quantum numbers, such as their data range, are surely another consequence of quantum measurement or observation. The information about quantum numbers are actually encoded in the effective evolution operator in Eq.~(\ref{unitary evolution operator})
\begin{equation}
\begin{aligned}
U_{\rm eff}(t) =&e^{-iH_{\rm eff}t}+\frac{1}{K}\left[\left(\sum_n\frac{Q_{n}}{a_{n}}HP_{n}\right)e^{-iH_{\rm eff}t}+e^{-iH_{\rm eff}t}\left(\sum_nP_{n}H^{\dagger}\frac{Q_{n}}{a_{n}}\right)\right]\\
=&e^{-iH_{\rm eff}t}+\frac{1}{K}\left[\left(\sum_n\sum_{m\neq n}\frac{P_{m}HP_{n}}{\eta_{n}-\eta_{m}}\right)e^{-iH_{\rm eff}t}+e^{-iH_{\rm eff}t}\left(\sum_n\sum_{m\neq n}\frac{P_{n}H^{\dagger}P_{m}}{\eta_{n}-\eta_{m}}\right)\right]\\
=&e^{-iH_{\rm eff}t}+\frac{1}{K}\left[\left(\sum_n\sum_{m\neq n}\frac{H_{mn}}{\eta_{n}-\eta_{m}}|m\rangle\langle n|\right)e^{-iH_{\rm eff}t}+e^{-iH_{\rm eff}t}\left(\sum_n\sum_{m\neq n}\frac{H^{\dagger}_{nm}}{\eta_{n}-\eta_{m}}|n\rangle\langle m|\right)\right]
\end{aligned}
\end{equation}
Recall this equation is a typical Taylor expression to the order of $\mathcal{O}(1/K)$. In our perturbative framework for the Zeno mechanism, it is consistent to require that
\begin{equation}
\frac{|H_{mn}|}{K|\eta_n-\eta_m|} \leq \mathcal{O}(K^0) \approx 1
\end{equation}
The same argument applies to the off-diagonal elements of $H^{\dagger}$. Note in the current section, we focus on the situation with non-vanishing $|H_{mn}|$. Then we have
\begin{equation}
|\eta_n-\eta_m|\geq \frac{|H_{mn}|}{K}
\end{equation}
In another word, the quantum numbers $\eta_n,\eta_m,\cdots$ evolved in the particle physics, such as electric charge and mass, are discretely spaced in magnitudes in the presence of a finite $K$.

A subsequent interesting phenomenon is the hierarchy pattern of mass. In our framework, the origin of masses of fundamental particles, such as charged leptons and quarks, is based on Zeno measurement mechanism. It is dramatically different from the Higgs mechanism in Standard Model, by which quarks and leptons acquire the mass. However, the masses in the context of Higgs mechanism are entirely arbitrary. The hierarchical pattern of lepton and quark masses then has to call for another theory~\cite{Weinberg:2020zba} and there have been many optional models to give these masses~\cite{Weinberg:2020zba, Weinberg:1972ws, S.M.Barr, F.Wilczek, T.Yanagida}. All of these studies are in the language of quantum field theory based on SM. But here we can interpret such a hierarchical pattern based on quantum measurement as an inspiring and heuristic work.

We recall the eigen-equation~(\ref{eigenfunction}), and the eigenvalues for measurement Hamiltonian $H_M$ are considered as the masses for the charged leptons (The similar argument also applies to quarks). Then we have
\begin{equation}
H_MP_{n}=m_{n}P_{n}, \quad P_n=|\l_n\rangle\langle\l_n|
\end{equation}
where $|l_n\rangle$ denotes the eigenstate of the $n$-th generation charged lepton with $l_n\in\{e,\mu,\tau\}$ (implying the leptons of electron, muon, tauon, respectively). The charged leptons are either in the mass representation, by which Eq.~(\ref{conservation2}) describes the condition for energy-momentum conservation or in the flavor representation, by which Eq.~(\ref{conservation2}) is related to the conservation of fermion number. The latter should also be satisfied for the sake of the electric-charge conservation. Due to the preceding argument, these two representations can be the same one. Otherwise someone of the two conservation laws will be broken. So that {\em Condition B} becomes the unique solution for the charged leptons. If the rest masses for the $3$-generation charged leptons can be ordered as $m_e<m_{\mu}<m_{\tau}$, then {\em Condition B} reads
\begin{equation}\label{lepton solution}
m_{\mu}-m_e\geq\mathcal{O}(K), \quad m_{\tau}-m_e\geq\mathcal{O}(K), \quad m_{\tau}-m_{\mu}\geq\mathcal{O}(K)
\end{equation}
which indicates very large mass gap among the three charged leptons. This results are consistent with the hierarchy pattern of the masses of charged leptons given by $m_e\approx5.1\times10^{5}$ eV, $m_{\mu}\approx1.06\times10^{8}$ eV, $m_{\tau}\approx1.8\times10^9$ eV~\cite{Zyla:2020zbs}. Similarly we can also find that the masses of different flavor of quarks should also have a hierarchical pattern.

It is emphasised that the double conservations (energy and charge conservations) in the examples of electric charged leptons or quarks give rise to a hierarchy pattern of masses and naturally rule out the possibility of oscillation or mixing in {\em Condition A}. Comparing the analysis in Sec.~\ref{neutrino} and the current section, we can also propose that the fermions carrying vanishing masses cannot be electrically charged.

\subsection{Color confinement}\label{confine}

Color-charged elementary particles are known as quarks and gluons in SM. The color-charged particles exchange gluons via strong interactions as the electrically-charged particles exchange photons via electrical-magnetic interactions. Each quark carries one of the three color charges and constantly changes its color charge when exchanging gluons with the other quarks. During the process that a quark emits or absorbs a gluon, the quark's color must change in order to preserve color charge, i.e., the color charge is always conserved. The color-charged particles cannot be found individually in experiments, and the color-charge quarks are confined in groups (hadrons) with other quarks. These composites are color neutral. Color confinement represents the most amazing phenomenon in quantum chromodynamics (QCD). Despite the tremendous progress in QCD since its invention, however, the origin and mechanism of color confinement remain unclear and open problems in all models~\cite{Nambu:1974zg,Mandelstam:1974pi, Ambjorn:1980ms,Engelhardt:1999fd,Greensite:2016pfc, Olesen:2016pxv}. Recently, several works are dedicated to study the confinement mechanism~\cite{Chen:2020hsl, Pak:2020izt, Pak:2020fkt} based on quantum field theory.

Once again, we now study the phenomenon of color confinement based on Zeno mechanism. We can rewrite Eq.~(\ref{eigenfunction}) with respect to the eigen-structure for the colored quarks
\begin{equation}
H_MP_n=c_nP_n, \quad P_n=|C_n\rangle\langle C_n|
\end{equation}
where $|C_1\rangle=|R\rangle$, $|C_2\rangle=|G\rangle$, and $|C_3\rangle=|B\rangle$ denote the eigenstate of color charge, respectively. And one can express the corresponding eigevalues as $c_1=r$, $c_2=g$, and $c_3=b$.

Beyond {\em Condition B}, the double conservations (energy-momentum conservation and color-charge conservation) leaves another solution: the dimension number of the color representation collapses to one as stated by {\em Condition C}, i.e., the color charges are confined or blind and then the color eigenstates have to be combined into one neutral state. The blind manner or mechanism for these color charges is not unique. A possible solution is that the color charge $|C_n\rangle$ and its anti-color charge (denoted by $|\bar{C_n}\rangle$) can be cancelled with each other and in this case the color-neutral composite state reads $|C_n\bar{C_n}\rangle$, such as meson including pion, kaon, etc. Another one is that all of three color charges or three anti-color charges are combined together, then the color-neutral composite state reads $|RGB\rangle$, such as baryon including proton, neutron, etc. We cannot list all of the color-neutral combinations allowed in principle, e.g., the pentaquark states~\cite{Zyla:2020zbs}.

In any manner of confinement, the color charges have to be close to each other in the distance of quarks, that is determined by the energy-scale $\Lambda$ of quarks and the coupling-constant $g_s$ of strong interaction. According to uncertainty principle, a larger $\Lambda$ means a shorter distance between (light) quarks in the relativistic limit as well as a stronger confinement and vice versa. On the other hand, a larger coupling-constant $g_s$ also provides a stronger attractive force linking the quarks. Note a very large $K$ that renders $H_{\rm eff}\approx KH_M$ determines a very small time-interval $t\sim\mathcal{O}(1/K)$. The mission of color confinement for the quarks combining into a neutral composites then should be ``completed'' in such a short time-scale. Otherwise there will be a chance to observe a single quark. We have $\Lambda\sim1/t\sim\mathcal{O}(K)$ by virtue of the time-energy uncertainty relation. Then if the quarks' energy-scale $\Lambda$ decreases by enlarging the distance between quarks, the time-scale for color confinement becomes longer and the probability to ``see'' a single quark will be enhanced. In this case, the coupling-constant $g_s$ of strong interaction have to play a key role to ensure the cancellation of different color-charges carried by quarks, i.e., $g_s$ will become larger to pull quarks back together to combine into a color-neutral composite. Therefore we can deduce that $g_s=g_s(\Lambda)$ is a monotonically decreasing function with respect to $\Lambda$, which indicates the so-called asymptotic-freedom phenomenon in QCD. Then asymptotic-freedom is closely associated with color confinement by Zeno mechanism.

In addition, we can show that the color confinement is not necessary to the color conservation if we only have two types of colors. Consider a model having two color charges $c_1$ and $c_2$ and the eigen-projectors are defined as $P_1=|C_1\rangle\langle C_1|$ and $P_2=|C_2\rangle\langle C_2|$. Then we have [see Eq.~(\ref{unitary evolution operator})]
\beq\label{twocolor}
\left(\sum_n\frac{Q_{n}}{a_{n}}HP_{n}\right)=\frac{P_2HP_{1}}{c_1-c_2}+\frac{P_1HP_2}{c_2-c_1}
=\frac{1}{c_1-c_2}(P_2HP_1-P_1HP_2)=\frac{1}{c_1-c_2}(P_2HP_1-P_2H^{\dagger}P_1)
\eeq
If the Hamiltonian $H$ is Hermitian, then Eq.~(\ref{twocolor}) as well as the whole transition amplitude vanishes, and then the evolution operator in Eq.~(\ref{unitary evolution operator}) becomes $K$-independent, i.e., $U_{\rm eff}(t)=e^{-iH_{\rm eff}t}$. The color-charge conservation is consequently held by Eq.~(\ref{conservation2}). We thus conjecture that for a Hermitian $H$, color confinement becomes necessary to the color-charge conservation when the particle contains more than two types of color charges. Yet if the Hamiltonian is not Hermitian $H\neq H^{\dagger}$, then the color confinement is always demanded.

\section{Discussion and outlook}\label{discussion}

In this work, we address the question, that is, whether the effective Hamiltonian or model as a result of quantum Zeno effect, may describe the eigenstates of the measurement Hamiltonian for the fundamental particles in the sense that they emerge from the physics of high energy-scale [implied by $K$ in Eq.~(\ref{eq:sys+meas})] beyond the reach of measurement device of a finite resolution. Here we establish the conditions (A, B, C proposed in Sec.~\ref{phenomena}) that are the possible solutions for the Zeno mechanism based on our theoretical assumptions (1 and 2 proposed in Sec.~\ref{model}) and then present several examples. The theoretical results obtained in these examples are in the spirit that low-energy physics presents the underlying complicated processes determined by $KH_M$ from the high-energy correction through quantum measurement and the charge-conservation requirement. We focus on several-dimensional systems in the current work, although our results can be generalized to more dimensions with necessary aids from more constrains on parameters following the ideas presented here. That is to say, any condition should give rise to a certain set (maybe more than one) of properties that is used to identify fundamental particles (maybe quasi-particles), which is subject to the confirmation by measurements.

In particular, the basic idea or the top-down strategy throughout this study is that any observable physics is the subject of measurement, and formally they can be described by the effective Hamiltonian projected in the Zeno subspace determined by a strong-coupling model. The fundamental particles that can be directly observed are nothing but quantum numbers in their eigen-representation. In physics, particles are generated by unknown complicated processes described by $KH_M$. Simply in regarding charge-conservation or symmetry, we assume they are the manifestations of a underlying quantum Zeno mechanism by such quantum numbers that are certain conserved charges. In the last section, the prerequisites of charge-conservation have been connected with certain important phenomena in particle physics, such as neutrino oscillations, mass hierarchy pattern of charged fermions, discretization of quantum numbers, and color confinement. Rather than strictly ``deriving'' these physical results through quantum Zeno effect, we actually probe or interpret them through reasonable solutions or conditions meeting the requirement by the vanishing transition-amplitudes in Eq.~(\ref{conservation2}) as obtained from the effective model.

Our model connects separable regions with respect to the energy scale. The weak-coupling region is identified as the physical world that makes sense to our measurement device with a finite resolution. The strong-coupling region can not be directly observed while its physical impact can be illustrated as quantum Zeno effect relevant to observation. What we indeed observe is the physical sectors projected by the projectors determined by the measurement we can perform. In this work, the quantum numbers that we have considered include mass, electric charge, and color charge. Thus it is far-off completing the whole story, but only the very beginning. Firstly, some other important quantum numbers, such as spin, linear momentum and angular momentum are yet not discussed by our mechanism. Secondly, the classical symmetry broken by the quantum corrections in quantum field theory, such as the well-known chiral anomaly, might be still hidden in the high-order terms in Eq.~(\ref{eq: neutrino}), such as $\mathcal{O}(1/K^2)$, that is to be considered on tuning the coupling strength $K$ from high-energy to low-energy scales. Thirdly, we can understand that the charged fermions, leptons and quarks, are discrete in mass via our model, yet a more detailed investigation to predict the exact values of masses of such fermions is still underway. Fourth, the present work focuses on particles but not include anti-particles. The famous question, that the amount of particles are far more than that of anti-particles, can be reasonably treated in two possible ways: (1) the mechanism for the production of particles and anti-particle are totally different from each other; or (2) the coupling constants $K$ for the production of the anti-particles are much weaker than those for particles.

Note the rotating-wave approximation, a technique popularly applied in the quantum optics and quantum information science, is necessary to obtain the correct formula for neutrino oscillation in appendix~\ref{useful equations}. This approximation ignores the high-frequency oscillation terms which hardly contribute to the observation of neutrino physics. Therefore, the physics at very high energy scale can also be omitted as well. Then comes an interesting conjecture: the neutrino oscillation that has been observed might indicate that there exist novel physics at a very high-energy scale unknown to us. The unknown physics at the very high-energy scale demonstrates the smallness of their masses the other way round. We have therefore proposed a closed logical chain: light masses of neutrinos $\Rightarrow$ neutrino oscillation by using rotating-wave approximation $\Rightarrow$ unknown physics at high energy-scale $\Rightarrow$ light masses of neutrinos.

\section{Conclusion}\label{conclusion}

This work is a first contribution to an uncompleted theoretical framework, specifically with regards to probing the fundamental particles and their physics as the results of quantum measurement. Currently the framework can be described in the language of Zeno effect or Zeno subspace and standard perturbation theory. And here it is a purely quantum mechanical model of observation through projection of a measurement Hamiltonian. The basic idea is that observation or measurement is the source of fundamental particles. They can be probed by an effective Hamiltonian constructed by the eigenstates of some conserved charge, that is directly observed by measurement devices. Given the two {\em assumptions} in Sec.~\ref{model}, the fundamental particle physics can be partially understood by the self-consistent solutions to the vanishing transition amplitude between arbitrary pair of eigenstates for the conserved charges. Due to the formation for the main results presented in Eq.~(\ref{Heff}) $\sim$ Eq.~(\ref{conservation2}), temporally we call our theoretical framework as Zeno mechanism.

\begin{table}[htbp]\centering
\begin{tabular}{|c|c|c|} \hline
Particles/Quantum number  & Physical prerequisite & Physical implications \\ \hline
Neutrinos & Energy-momentum conservation & Small mass; Oscillations; Big-angle mixing  \\ \hline
Electric-charged fermions & Electric-charge conservation & Hierarchical pattern of masses \\ \hline
All fundamental quantum number & Finite measurement strength $K$ & Discretization (Quantization?) \\ \hline
Color-charged fermions & Color charge conservation & Color confinement \\ \hline
\end{tabular}
\caption{The particle physics that have been covered by Zeno mechanism in this work. }
\label{conctable}
\end{table}

In Table~\ref{conctable}, we summarize the phenomena and physical conditions that have been discussed under the Zeno mechanism formally by Eq.~(\ref{conservation2}) embodying various physics. To preserve the energy-momentum conservation, we show that for the very light neutrinos, the emergence of the neutrino-oscillations in the flavor representation and the mixing angles for neutrinos cannot be very small. To preserve the fermion number, we show that the mass gap should be larger than the energy-scale of the measurement Hamiltonian, that indicates the hierarchical pattern of masses for electric-charged fermions. To work with a finite energy-scale of the measurement Hamiltonian and the consistency for a perturbative model, all fundamental quantum numbers are required to be discrete in magnitude. To contract the number of the degrees of freedom involved in Eq.~(\ref{conservation2}), color confinement can also be induced by the vanishing transition amplitudes.

Our Zeno mechanism is established to separate the conceived ``physical-system Hamiltonian'' and the observable information or structure obtained by quantum measurement, that makes sense to the fundamental postulation for the particle physics. Through enlarging the size of the Zeno subspace and regarding higher energy-scale, we are expecting to discover more interesting and novel physics.

\section*{Acknowledgments}

We acknowledge grant support from the National Science Foundation of China (Grants Nos. 11974311 and U1801661), Zhejiang Provincial Natural Science Foundation of China under Grant No. LD18A040001, and the Fundamental Research Funds for the Central Universities (No. 2018QNA3004).

\appendix

\section{Probability formula for the light neutral fermion (neutrino) oscillation}\label{oscillationFormula}

This appendix is contributed to deducing the probability formula for the oscillation of the very light neutral fermions (neutrinos) based on the energy-momentum conservation. The formula is shown to be a consequence under {\em Condition A} proposed in Sec.~\ref{phenomena} for the Zeno mechanism. In mathematics, it is then demanded that Eq.~(\ref{conservation2}) must be valid, i.e., the $K$-dependent term in Eq.~(\ref{transitionamplitude}) or Eq.~(\ref{unitary evolution operator}) that break the energy-momentum conservation must be cancelled or suppressed. The derivation is given in appendix~\ref{useful equations}. With this result, the transition amplitude $\mathcal{A}(\nu_{\alpha}\to\nu_{\beta})$ follows the time evolution purely governed by the effective Hamiltonian $H_{\rm eff}$ in Eq.~(\ref{Heff}). In the interaction picture with respect to $H_M$, it reads
\begin{equation}\label{ori-transition}
\mathcal{A}(\nu_{\alpha}\to\nu_{\beta})=\langle\nu_{\beta}|e^{-i(\sum_{i}P_iHP_i)t}|\nu_{\alpha}\rangle
\end{equation}
where $|\nu_{\alpha}\rangle$ and $|\nu_{\beta}\rangle$ are the eigenstates of the light neutral fermions (neutrinos) in the flavor representation. Note in the mass representation, neutrinos will merely accumulate dynamical phases according to their various masses when propagating in space. It is very natural and necessary to express Eq.~(\ref{ori-transition}) in the mass representation by using the transformation matrix connecting these two representations as given by Eq.~(\ref{u-matrix}). Then we have
\begin{equation}\label{right}
\mathcal{A}\left(\nu_{\alpha}\to \nu_{\beta}\right)=\sum_{k,l}\langle\nu_{\beta}|\nu_k\rangle\langle \nu_k|e^{-i(\sum_{i} P_{i} H P_{i})t }|\nu_l\rangle\langle\nu_l|\nu_{\alpha}\rangle=\sum_{k,l}V^{*}_{\beta k}V_{\alpha l} =\sum_{i}V^{*}_{\beta i}V_{\alpha i}e^{-iE_it}
\end{equation}
where $E_i\equiv\langle\nu_i|H|\nu_i\rangle$. Consequently, the transition probability reads
\begin{equation}\label{generaltransition}
P(\nu_{\alpha}\to \nu_{\beta})=|\mathcal{A}(\nu_{\alpha}\to  \nu_{\beta})|^{2}=\sum_{i}|V_{\alpha i}|^{2}|V_{\beta i}|^{2}+\sum_{i>j}(V_{\alpha i}V^{*}_{\beta i}V^{*}_{\alpha j}V_{\beta j})e^{-i(E_{i}-E_{j})t}+(i \leftrightarrow j)
\end{equation}
In the extremely relativistic limit for very light neutral fermion (neutrinos), all the neutral fermions (neutrinos) carry the same mean energy $E$ and momentum $p$ with $E\approx p\gg m_i$, then $E_i$'s and their gaps can be expressed by
\begin{equation}\label{relativistic limit}
E_{i}=\sqrt{p_{i}^{2}+m_{i}^{2}}\simeq p+\frac{m_{i}^{2}}{2 E}, \quad E_{i}-E_{j}\simeq\frac{m_{i}^{2}-m_{j}^{2}}{2E}\equiv \frac{\Delta m_{ij}^{2}}{2E}
\end{equation}
respectively. Subsequently, Eq.~(\ref{generaltransition}) can be written as
\begin{equation}\label{generaltransition1}
P(\nu_{\alpha}\to  \nu_{\beta})=\sum_{i}|V_{\alpha i}|^{2}|V_{\beta i}|^{2}+\sum_{i>j}(V_{\alpha i}V^{*}_{\beta i} V^{*}_{\alpha j}V_{\beta j})\exp\left(-i\frac{\Delta m_{ij}^{2}}{2E}L\right)+(i\leftrightarrow j)
\end{equation}
Note here it is supposed that the neutrino is detected along its direction of motion after it travels a distance $L$ with the mean energy $E$ and in the natural units, $\hbar=c=1$, $L\approx ct\equiv t$.

One can rewrite the elements of unitary transformation matrix in Eq.~(\ref{generaltransition1}) as
\begin{equation}\label{u-para}
V_{\alpha i}V^{*}_{\beta i}V^{*}_{\alpha j}V_{\beta j}=\left|V_{\alpha i} V^{*}_{\beta i} V^{*}_{\alpha j}V_{\beta j}\right| \exp\left(i\phi_{\alpha\beta;ij}\right)
\end{equation}
where the phase $\phi_{\alpha\beta;ij}={\rm Arg}(V_{\alpha i}V^{*}_{\beta i}V^{*}_{\alpha j}V_{\beta j})$ satisfying $\phi_{\alpha\beta;ij}=-\phi_{\alpha\beta;ji}$. Inserting Eq.~(\ref{u-para}) into Eq.~(\ref{generaltransition1}), one can find
\begin{equation}\label{preprobability}
P(\nu_{\alpha}\to \nu_{\beta})=\sum_{i}\left|V_{\alpha i}\right|^{2}\left|V_{\beta i}\right|^{2}+2\sum_{i>j}\left|V_{\alpha i}V^{*}_{\beta i}V^{*}_{\alpha j}V_{\beta j}\right|\cos\left(\frac{\Delta m_{i j}^{2}}{2E}L-\phi_{\alpha\beta;ij}\right)
\end{equation}
Considering the case where the final state is the same as the initial state $\beta=\alpha$, it is straightforward to see that
\begin{equation}\label{result}
P(\nu_{\alpha}\to \nu_{\beta})=\delta_{\alpha\beta}-4\sum_{i>j}\Re\left(V_{\alpha i}V^{*}_{\beta i} V^{*}_{\alpha j} V_{\beta j}\right)\sin^{2}\left(\frac{\Delta m_{ij}^{2}L}{4E}\right)+2\sum_{i>j}\Im\left(V_{\alpha i}V^{*}_{\beta i} V^{*}_{\alpha j} V_{\beta j}\right)\sin\left(\frac{\Delta m_{ij}^{2}L}{2E}\right)
\end{equation}
where $\Re(\cdot)$ and $\Im(\cdot)$ imply the real and imaginary parts of a complex number, respectively.

\section{Solution for neutral fermions (neutrinos)}\label{useful equations}

In this appendix, we apply the Zeno mechanism for the physics of light neutral fermions under {\em Condition A} to obtain a consistent solution implying the phenomenon of neutrino oscillation together with big-angle mixing. The derivation to obtain a $K$-independent transition amplitude $\mathcal{A}(\nu_{\alpha}\to\nu_{\beta})$ with $\alpha,\beta\in\{e,\mu,\tau\}$ in the flavor representation is focused on the three-generation neutrinos, which constitutes a sufficient condition to meet the requirement of vanishing transition amplitude $\mathcal{A}(m_k\to m_n)$ with $k,n\in\{1,2,3\}$ in the mass representation by imposing the energy-moment conservation.

To get the consistent solutions, now we calculate the $\mathcal{O}(1/K_m)$ terms of transition amplitude $\mathcal{A}(\nu_{e}\to\nu_{\mu})$ as an example, which can be denoted by $\mathcal{A}_{vio}(\nu_{e}\to\nu_{\mu})$. Using Eq.~(\ref{unitary evolution operator}), we have
\begin{equation}\label{example}
\begin{aligned}
\mathcal{A}_{vio}(\nu_{e}\to\nu_{\mu})&=\langle\nu_{\mu}|\frac{1}{K}\left[\left(\sum_n\frac{Q_{n}}{a_{n}}HP_{n}\right) e^{-iH_{\rm eff}t}+e^{-iH_{\rm eff}t}\left(\sum_nP_{n}H^{\dagger}\frac{Q_{n}}{a_{n}}\right)\right]|\nu_e\rangle \\
=&\sum_{i<j}\frac{V_{ei}V^{*}_{\mu j}}{K_m m_{ji}}\left[\langle\nu_{i}|H|\nu_{j}\rangle e^{-i(K_m m_j+E_j)t}-\langle \nu_{i}|H^{\dagger}|\nu_{j}\rangle e^{-i(K_m m_i+E_i)t}\right] \\
&+\sum_{i<j}\frac{V_{ej}V^{*}_{\mu i}}{K_m m_{ij}}\left[\langle\nu_{j}|H|\nu_{i}\rangle e^{-i(K_m m_i+E_i)t}-\langle \nu_{j}|H^{\dagger}|\nu_{i}\rangle e^{-i(K_m m_j+E_j)t}\right] \\
=&\frac{1}{K_m}\sum_{i<j}e^{-i(K_m m_i+E_i)t}\left(\frac{V_{ej}V^{*}_{\mu i}}{m_{ij}}\langle\nu_{j}|H|\nu_{i}\rangle-\frac{V_{ei}V^{*}_{\mu j}}{m_{ji}}\langle\nu_{i}|H^{\dagger}|\nu_{j}\rangle\right)\\ &+\frac{1}{K_m}\sum_{i<j}e^{-i(K_m m_j+E_j)t}\left(\frac{V_{ei}V^{*}_{\mu j}}{m_{ji}}\langle\nu_{i}|H|\nu_{j}\rangle-\frac{V_{ej}V^{*}_{\mu i}}{m_{ij}}\langle \nu_{j}|H^{\dagger}|\nu_{i}\rangle\right)\\
=&\frac{1}{K_m}\left[e^{-i(K_m m_1+E_1)t}(X_{21}+X_{31})+e^{-i(K_m m_2+E_2)t}(X_{12}+X_{32})+e^{-i(K_m m_3+E_3)t}(X_{13} +X_{23})\right]\\
\approx& \frac{1}{K_m}e^{-iEt}\left[e^{-iK_m m_1t}(X_{21}+X_{31})+e^{-i K_m m_2t}(X_{12}+X_{32})+e^{-i K_m m_3t}(X_{13} +X_{23})\right]
\end{aligned}
\end{equation}
where
\begin{equation}\label{parameter}
X_{ij}\equiv H_{ij}\frac{V_{ei}V^{*}_{\mu j}}{m_{ji}}+H_{ji}^{\dagger}\frac{V_{ej}V^{*}_{\mu i}}{m_{ji}}
\end{equation}
with $m_{ij}\equiv m_i-m_j$ and $H_{ij}=\langle\nu_{i}|H|\nu_{j}\rangle$, and $H_{ij}^{\dagger}=\langle\nu_{j}|H|\nu_{i}\rangle^{\dagger}$. Note in the last line of Eq.~(\ref{example}), we apply the extremely relativistic approximation $E_i\approx E$ by using Eq.~(\ref{relativistic limit}) and drop the term $\mathcal{O}(m_i^2/(2E))$ for $m_i^2/(2E)\ll K_m m_i$. Similarly, we can obtain another two expressions for $\mathcal{A}_{vio}(\nu_{\tau}\to\nu_{e})$ and $\mathcal{A}_{vio}(\nu_{\mu}\to\nu_{\tau})$ by the definitions: 
\begin{equation}
Y_{ij}=H_{ij}\frac{V_{ei}V^{*}_{\tau j}}{m_{ji}}+H_{ji}^{\dagger}\frac{V_{ej}V^{*}_{\tau i}}{m_{ji}}, \quad
Z_{ij}=H_{ij}\frac{V_{\mu i}V^{*}_{\tau j}}{m_{ji}}+H_{ji}^{\dagger}\frac{V_{\mu j}V^{*}_{\tau i}}{m_{ji}}
\end{equation}

To have a nontrivial solution for Eq.~(\ref{example}) under arbitrary combinations of particle masses and transformation matrix, we can set one of the masses $m_i\approx0$ ($i\in\{1,2,3\}$) and then apply the rotating wave approximation (RWA) to the remaining high-frequency terms of $e^{-iK_m m_kt}$ and $e^{-iK_m m_nt}$ with $i\neq k,n$ and $k\neq n$. Consequently, we have
\begin{equation}\label{XYZ}
X_{ki}+X_{ni}=0, \quad Y_{ki}+Y_{ni}=0, \quad Z_{ki}+Z_{ni}=0
\end{equation}
These algebraic equations always have nontrivial solutions since they are equivalent to a group of $6$ linear equations for $4$ real variables determined by $H_{ki}$, $H_{ik}^{\dagger}$, $H_{ni}$, and $H_{in}^{\dagger}$. It is remarkable to find that there are always nontrivial solutions for the $N$-generation ($N>3$) neutral fermions if one of the masses $m_i$ ($i\in\{1,2,\cdots,N\}$) is nearly zero and all the other nonvanishing masses generate high-frequency oscillations that can be omitted by RWA. Our model thus allows to be generalized to more generation neutral fermions.

Now we have cancelled all of the $\mathcal{O}(1/K_m)$ terms Eq.~(\ref{transitionamplitude}) through Eq.~(\ref{example}). Then we have
\begin{equation}\label{generalamp}
\mathcal{A}(\nu_{\alpha}\to \nu_{\beta})=\sum_{i}V_{\alpha i}V^{*}_{\beta i}e^{-i\frac{m_{i}^{2}t}{2E}}
\end{equation}
It is important to check the energy-momentum conservation as required by Eq.~(\ref{conservation2}). Under {\em Condition A}, we have
\begin{equation}\label{transition1}
\begin{aligned}
\mathcal{A}(\eta_i\to\eta_j)=&\mathcal{A}(m_i\to m_j)=\sum_{\alpha,\beta}V_{\beta j}^{*}V_{\alpha i}\mathcal{A}(\nu_{\alpha}\to\nu_{\beta})\\ =&\sum_{\alpha,\beta}V_{\beta j}V^{*}_{\alpha i}\sum_{l} V_{\alpha l} V^{*}_{\beta l} e^{-i\frac{m_{l}^{2}t}{2 E}}\\ =&\sum_{l}\delta_{li}\delta_{jl}e^{-i\frac{m_{l}^{2}t}{2 E}}\\ =& \delta_{ij}e^{-i\frac{m_{j}^{2}t}{2 E}}
\end{aligned}
\end{equation}
where we have used the orthonormality for the unitary matrix $\sum_{\alpha}V_{\alpha l}V^{*}_{\alpha i}=\delta_{li}$. Then Eq.~(\ref{conservation2}) is recovered and energy-momentum conservation is preserved.

Next our model allows an analysis in quality for the mixing angles. Under the assumption that the mixing angles are small, the transformation matrix in Eq.~(\ref{V-matrix}) can be expressed as
\begin{equation}\label{small}
V=\left(\begin{array}{ccc}
1 & \delta \theta &  \delta \theta \\
 \delta \theta &1 &  \delta \theta \\
 \delta \theta &  \delta \theta & 1
\end{array}\right)
+ \mathcal{O}(\delta\theta^2)
\end{equation}
Inserting the matrix elements in Eq.~(\ref{small}) to Eq.~(\ref{XYZ}), we find (in the case with $m_1\approx0$ and $m_2,m_3>0$) that
\begin{equation}
\begin{aligned}\label{mixingangle}
\frac{H_{21}\delta\theta^2}{m_{12}}+\frac{H_{12}^{\dagger}}{m_{12}}+\frac{H_{31} \delta\theta^2}{m_{13}}+\frac{H_{13}^{\dagger}\delta\theta}{m_{13}}=0\\
\frac{H_{21}\delta\theta^2}{m_{12}} +\frac{H_{12}^{\dagger}\delta\theta}{m_{12}} +\frac{H_{31}\delta\theta^2}{m_{13}}+\frac{H_{13}^{\dagger}}{m_{13}}=0\\
\frac{H_{21}\delta\theta}{m_{12}}+\frac{H_{12}^{\dagger} \delta\theta^2}{m_{12}}+\frac{H_{31}\delta\theta^2}{m_{13}}  +\frac{H_{13}^{\dagger}\delta\theta}{m_{13}}=0
\end{aligned}
\end{equation}
It turns out that the magnitude of $H_{21}$, $H_{31}$, $H_{12}^{\dagger}$, and $H_{13}^{\dagger}$ are in the order of $\mathcal{O}(\delta\theta^2)$ for small $\delta\theta$. The small magnitudes of the Hamiltonian corresponds to the seemingly trivial solution discussed in Sec.~\ref{phenomena} or the condition in which the mass representation is nearly the same as the flavor representation. Similar analysis applies to the other two conditions with $m_2\approx0$ or $m_3\approx0$. Therefore in a great probability, the mixing angles among the flavor eigenstates can not be too small.

Generally, one can write the transformation matrix $V$ as~\cite{Akhmedov:1999uz}
\begin{equation}\label{V-matrix}
V=\left(\begin{array}{ccc}
c_{12} c_{13} & s_{12} c_{13} & s_{13} e^{-i\gamma} \\
-s_{12} c_{23}-c_{12} s_{23} s_{13} e^{i\gamma} & c_{12} c_{23}-s_{12} s_{23} s_{13} e^{i\gamma} & s_{23} c_{13} \\
s_{12} s_{23}-c_{12} c_{23} s_{13} e^{i\gamma} & -c_{12} s_{23}-s_{12} c_{23} s_{13} e^{i\gamma} & c_{23} c_{13}
\end{array}\right)
\end{equation}
with $c_{ij}\equiv\cos\theta_{ij}, s_{ij}\equiv\sin\theta_{ij}$. According to experimental data for neutrino physics~\cite{SinghKoranga:2020jzh}: $\Delta m_{21}^{2}=7.55\times10^{-5} {\rm eV}^2$, $|\Delta m_{31}^{2}|=2.50\times10^{-3} {\rm eV}^2$, $\sin^{2}\theta_{12}=0.32$, $\sin^{2}\theta_{23}=0.547$, $\sin^{2}\theta_{13}=0.0216$, and $\gamma=1.32\pi$. Then the transformation matrix in Eq.~(\ref{V-matrix}) becomes
\begin{equation}\label{data matrix}
V=\left(\begin{array}{ccc}
 0.82373 & 0.565074 & -0.0787501+0.12409 i \\
 -0.332708+0.0756809 i & 0.587961+0.0519167 i & 0.738795 \\
 0.462085+0.0688718 i & -0.579902+0.0472457 i & 0.672325
\end{array}\right)
\end{equation}
by which our model gives two results for the neutrinos' masses: (1) $m_1\approx0$ eV, $m_2\approx0.009$ eV, $m_3\approx0.050$ eV, and then $\sum_im_i\approx 0.059$ eV; (2) $m_3\approx0$ eV, $m_1\approx0.050$ eV, $m_2\approx0.051$ eV, and then $\sum_im_i\approx0.101$ eV. Note the possibility of $m_2\approx0$ has been ruled out by the experimental data.

\end{document}